# Development of an interface for digital neuromorphic hardware based on an FPGA

René Harmann[1,*], Lukas Sohlbach[1,2], Fernando Perez-Peña[2], and Karsten Schmidt[1]

[1] Department of Computer Science and Engineering, Frankfurt University of Applied Sciences, Nibelungenplatz 1, 60318 Frankfurt, Germany
[2] Department of Automation, Electronics and Computer Architecture and Networks, School of Engineering, University of Cadiz, Avda. de la Universidad 10, 11519 Puerto Real, Spain
harmann@stud.fra-uas.de (FA, CA)

**Abstract**
Exploring and understanding the functioning of the human brain is one of the greatest challenges for current research. Neuromorphic engineering tries to address this challenge by abstracting biological mechanisms and translating them into technology. Via the abstraction process and experiments with the resulting technical system, an attempt is made to obtain information about the biological counterpart. One subsection of Neuromorphic Engineering (NE) are Spiking Neural Networks (SNN), which describe the structures of the human brain more and more closely than Artificial Neural Networks (ANN). Together with their dedicated hardware, SNNs provide a good platform for developing new algorithms for information processing. In the context of these neuromorphic hardware platforms, this paper aims to develop an interface for a digital hardware platform (SPINN-3 Development Board) to enable the use of industrial or conventional sensors and thus create new approaches for experimental research. The basis for this endeavor is a Field Programmable Gate Array (FPGA), which is placed as a gateway between the sensors and the neuromorphic hardware. Overall, the developed system provides a robust solution for a wide variety of investigations related to neuromorphic hardware and SNNs. Furthermore, the solution also offers suitable possibilities to monitor all processes within the system in order to obtain suitable measurements, which can be examined in search of meaningful results.

# 1    Introduction

With the development of increasingly powerful computers and sophisticated sensor systems, machine learning algorithms have become more and more powerful. However, there is still a large gap between the human brain and current technology, which can be illustrated by the fact that a hypothetical clock-based computer at the level of the human brain would consume 12 GW of energy, compared to the 20 W of the human brain [1]. Thus, exploring and understanding the mechanics of the human brain is one of the greatest challenges for current research. Neuromorphic Engineering (NE) tries to address this challenge by abstracting biological mechanisms and translating them into technology. The NE field is based on Spiking Neural Networks (SNNs), which describe the structures and the event-based information processing of the human brain better, in a more detailed way than Artificial Neural Networks (ANN). To compute SNNs of different sizes in real-time various hardware solutions have been developed. Contrary to conventional chip designs where all computations are triggered by a global clock, in SNN chips calculations are only performed when an event signal, called spike, is present. Thus, this asynchronous model is more suitable for SNNs inherent sparseness [2]. In the context of these neuromorphic hardware platforms, this paper presents an interface between a SpiNNaker [3] development board (SPINN-3 [4]) and non-spiking sensors on which further work can be developed. Since most neuromorphic applications work with such sensors [1] and usually use an application-specific interface, a generic solution is introduced here. In the past, attempts have been made to standardize such an interface. In [5], a PC was used to convert protocols between SpiNNaker and various peripherals. However, this solution is not particularly well suited for mobile or portable applications [2]. To address this problem, [5] and [6] developed a Field Programmable Gate Array (FPGA) based interface which, however, only allowed unidirectional communication, namely sending data to the SpiNNaker. In [7], the authors presented a bidirectional approach by developing their interface based on a Complex Programmable Logic Device (CPLD). However, the capacity for external devices and sensors of this solution is limited to five ports [2]. To increase the number of ports an MCU interface was developed in [8]. Unfortunately, it does not have the capability for parallel data processing. Consequently, a real-time interface based on an FPGA is presented here. It is intended to expand the work of [5] and [6] and enable bidirectional communication. Therefore, some basic information about SNNs, especially about hardware platforms, specific protocols and coding strategies is presented in Section 2. Afterwards, the methodology used for the development is described in Section 3, before the actual hardware and software and simulation results are presented in Section 4. The paper finishes with a discussion and conclusion in Sections 5 and 6.

## 2 Spiking Neural Networks

SNNs are described as the third generation of Neural Networks (NN) [9]. Unlike conventional NNs, the behavior of spiking neurons depends not only on a threshold but also on the dynamics of a reactive membrane that is excited by synaptic inputs from presynaptic neurons. If the activation function of a traditional neuron reaches the defined threshold, the neuron will generate an output. A spiking neuron also generates an output once the threshold of the membrane potential is exceeded, but by considering the dynamics of the membrane the value of the output is no longer most important, but its temporal behavior. Thus, intermediate states between the two closed states firing and inactive can now also be displayed, which represents a considerable gain in information. Fig. 1 shows an example curve of a spiking neuron. Furthermore, all information about the dynamic behavior of the membrane potential is also being displayed. [10–12] The following illustration is inspired by the work of [11].

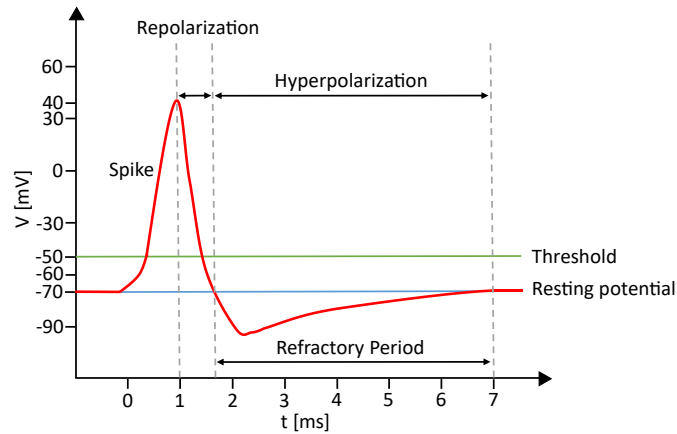

**Fig. 1.** Exemplary representation of a possible spike pattern with the additional representation of the different states of the membrane potential of a spiking neuron.

In addition to the information representation, SNNs also differ from conventional NNs in terms of the network structure. The neurons are not arranged in conventional layers but grouped together in so-called populations. Neurons in a population usually have a similar task. Populations of neurons can include recurrent connectivity as far as connections with other populations to build a complex structure. These additional degrees of freedom allow considerably more complex processing of the common information within the population. In order to allow external connectivity, each neuron is associated with a unique number within the population, which is considered as an address. The changed information representation and network structure also lead to different coding procedures: spike encoding and neuron encoding. [13–15]

In order to provide suitable hardware for SNNs, two approaches exist.

One is described as digital hardware, in which SNNs are mapped to dedicated hardware in the form of software. This creates an extremely high degree of flexibility in the design of the networks and changes can be easily applied. [16, 17]

However, the networks are only implemented digitally and do not represent a physical variant, which is the better approach for the approximation to the human brain. Nevertheless, there is also analog hardware that closes this gap. However, it should be noted that the implementation of the networks and the adaptability is much more difficult than on the digital platform. [16, 17]

## 3    Methodology

In this paper, only open source software solutions were used for the development of the interface. The FPGA was programmed in the language Verilog using the open source development suite Vivado 2022.2 [18]. Due to its extensive range of functions, Vivado allows the programming of all software components of the FPGA on different abstraction levels. In addition, the monitoring of internal signals and power consumption during runtime is feasible. The programming of the neuromorphic development board was done in Python, using a dedicated package called sPyNNaker [19] (version 1.6.0.0) and Visual Studio Code 1.78.2 as the development environment.

The hardware components used were the Nexys-A7 100T development board from Digilent for the FPGA and the development board SpiNN-3 from the University of Manchester for the neuromorphic hardware, which can be used to digitally map SNNs. For both development boards, the size and the available I/O options were important, because the interface should be applicable for variable use and the possibility to realize different protocols on the side of the FPGA had to be available. The SpiNN-3 development board uses four interconnected SpiNNaker cores.

In this approach, the Address Event Representation (AER) [20] is used for neuron information transfer between the two devices. Using this protocol, it is possible to design the communication between neurons so that the spike data can travel from the firing neuron to the receiving neuron. Thus, this protocol also allows the integration of external devices as independent neurons into the system. The actual exchange of information in this protocol is based on the addresses of the sender/receiver neurons. Furthermore, the rate coding method was selected, which allows the actual information content to be influenced by varying the transmission frequency. In this way it is possible to represent different resolutions of the information content, since no direct use of discrete values, except of the neuron addresses, is necessary and the information content lies only in the frequency.

For the implementation of the AER communication, the SpiNN-3 development board has two SpiNNaker link connectors which can establish bidirectional communication based on the SpiNNaker link protocol [21]. The SpiNNaker link protocol is used for modular communication between several external devices. On this basis, it has different routing keys that can identify the connected devices and thus enable error-free communication between them.

Within a data packet of the SpiNNaker link protocol the information content of the AER protocol is included. The AER data is integrated into the SpiNNaker link package using so-called mapper functions.

Thus, the SpiNNaker link protocol represents a combination of the communication between different hardware components and the neuron information.

Therefore, a special connector was established that uses high-speed level shifters to compensate for the different voltage levels of the FPGA and the SpiNN-3. Related to the circuits proposed to be implemented on the FPGA, some modules from the University of Manchester [22] (revision 2644) were used to establish the AER and SpiNNaker link communication. These modules were integrated into the general variable software design of the FPGA and can be accessed by any protocol.

In addition, the interface was equipped with a PC interface through which all data can be transferred to the target system during runtime. This enables further monitoring and data logging. The communication is based on the USB-UART standard. In addition, the communication setup between the FPGA and the neuromorphic hardware was established bidirectionally. This means that data from the SpiNN-3 can also be received and processed again.

## 4    Results

At the beginning of this chapter, the implementation of the physical connections of the interface is presented and afterwards the software and measurement results are addressed.

### 4.1    FPGA and SPiNN-3 connector

The physical connection between the FPGA and SpiNN-3 had to be specially developed, since no standardized connectors or cables exist for such a use case. The SpiNN-3 uses a 2-row 34-pin connector as input and output connector. For this reason, only a ribbon cable was suitable for the connection. Since the connectors of the SpiNN-3 are standardized, only the connector of the FPGA had to be developed. To save space, the connector was designed in a very small form factor. Voltage conversion is essential to establish a correct operation of the connection. The SpiNN-3 operates at 1.8 V on its connectors and the FPGA at 3.3 V. Two units of the Onsemi NLSV8T244 [23] were used as voltage converters. They were suited to achieve the desired transfer speeds with their maximum output delay of 3 ns specified by the manufacturer. Fig. 2 shows the exact circuit diagram of the signal paths on the designed connector from both devices.

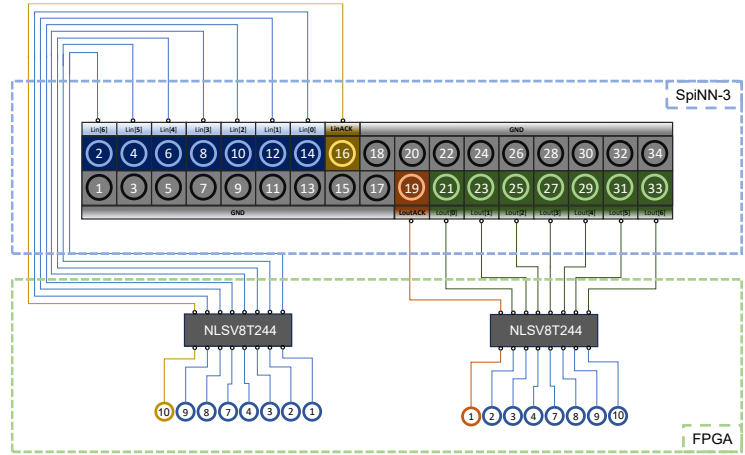

**Fig. 2.** Circuit diagram of the developed connector for the communication between FPGA and SpiNN-3

### 4.2 Software implementation

The structure of the software interface has a modular design as it can be seen in Fig. 3. Thus, the interface can be used with all available protocols and can process any sensor data. The structural design starts with reading the data of the selected sensor. Then the received data is converted into a frequency, which serves as the basis for the rate coding. With this frequency, the corresponding address of the neuron that is to be addressed on the SpiNN-3 is mapped into the AER protocol and subsequently into a SpiNNaker link packet and finally transmitted to the neuromorphic chip via the SpiNNaker link protocol. In detail, the modules were adapted to produce a pass-through mechanism to make them non-application specific. For this purpose, the AER mapper and the routing keys used were specially adapted. This means that only the address of the neuron is sent and no further information, such as the resolution of the sensor or similar. The response and the address with which the SpiNN-3 answers are decoded in the same way. As a basis for the communication on the side of the neuromorphic chip the class "external_device" and its function "activate_live_output_to" were used to receive and send data from the FPGA. It is important to note that the external device class must have the same characteristics (routing key, etc.) as they are set on the FPGA side. The information content can then be determined from the received packets, their transmission frequency and the neuron address. Subsequently the decoded information is further processed or transmitted to the device to be controlled. In addition to the actual tasks of the interface, it is included a visual feedback for the user in the form of a seven-segment display. This provides all the necessary information to the user and can be switched variably.

All interface data is also transmitted to the user via the PC connection. Since the PC is the main control unit of the interface, data such as a set value for a desired control can of course also be transmitted to the FPGA. Fig. 3 shows a graphical overview of the communication paths inside the interface.

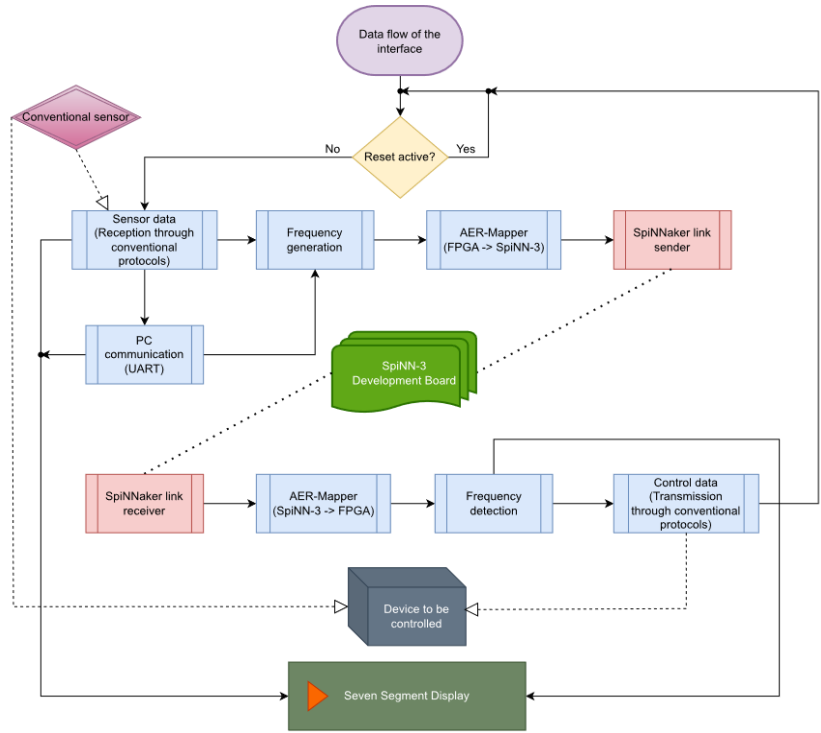

**Fig. 3.** Representation of the entire data flow of the developed interface between the software and hardware components used

### 4.3 Measurements

An experiment to check the bidirectional communication between the FPGA and the SpiNN-3 was performed. At the beginning, the neuron address 0 was sent to the SpiNN-3 with a frequency of 1 Hz and afterwards the frequency was increased to 10 Hz. Inside the SNN, there was a projection from the external device population to a second population named set value created. The output of the set value population was then connected and sent back to the external device population using the live output function. For the set value population, the IF_curr_exp neuron model and for the projection a static synapse was used, utilizing the parameters from Table 1.

**Table 1.** SNN parameters.

| $\tau_m$ [ms] | $\tau_{syn_E}$ [ms] | $\tau_{syn_I}$ [ms] | $\tau_{refrac}$ [ms] | $weight$ [-] | delay [ms] |
|---|---|---|---|---|---|
| 3 | 0.5 | 0.5 | 0.0 | 40.9 | 0.0 |

In Fig. 4 the spikes of the membrane potential and the spikes of the set value population are displayed. For the simulation a time step of 1000 µs and a time scale factor of 1 was used.

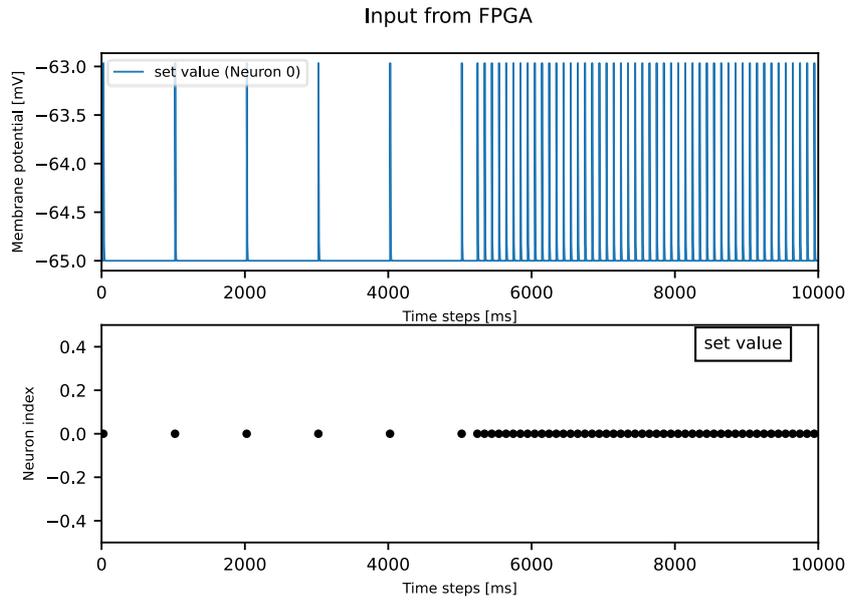

**Fig. 4.** Recorded neuron response on the SpiNN-3 excited by spikes send from the FPGA

The upper part of Fig. 4 shows a fast and short response of the membrane potential to the current injected from the external device population. When the threshold value of -50 mV is reached, a spike is fired, and the membrane potential is reset to the resting value of -65 mV. In the lower part of Fig. 4, a uniform distribution of spikes can be seen, and it is clearly visible that the spiking frequency of the population increases with the transmission frequency. In the simulation time of 10 s, 54 spikes occur.

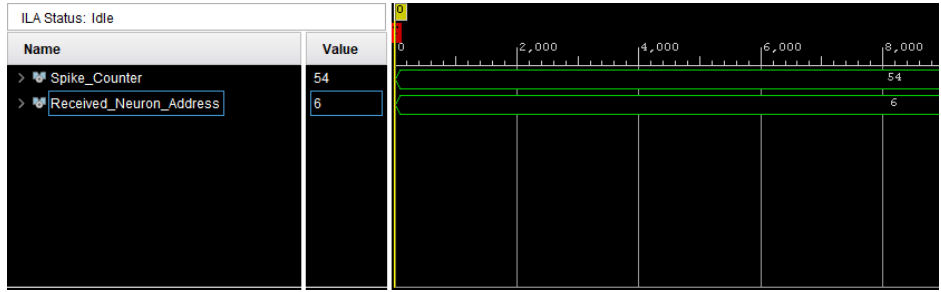

**Fig. 5.** Recorded answer from the SpiNN-3 development board received on the FPGA

Fig. 5 shows a screenshot of the Integrated Logic Analyzer (ILA) of Vivado. A spike of the neuron 0 of the set value population is coded with the address 6 due to the configuration of the network and the simulation software. The received address can be seen in the lower line of the ILA scope. In the upper line the final value of a counter is shown, which is incremented when address 6 is received. The value corresponds to the expected number of spikes.

In addition to the general consideration of the communication results, the power consumption of the interface was also important since one of the objective of the paper was to achieve an efficient solution in terms of power consumption. Using the power report provided by Vivado, the entire interface including its periphery shows a total power consumption of 242 mW. The biggest part of the total consumption is taken up by the mixed-mode clock manager.

## 5     Discussion

In [4] it is mentioned that the ports of the SpiNN-3 are 3.3 V sensitive. Thus, to detect a high level generated by the FPGA, for example, the level shifters mentioned in Fig. 2 are not really necessary. However, since a handshake between FPGA and SpiNN-3 is necessary for continuous communication via the SpiNNaker link protocol, the signals of the SpiNN-3 (pins 19-33 in Fig. 2) must be raised from 1.8 V to 3.3 V, otherwise they cannot be recognized by the FPGA and the communication is aborted after the first transmission attempt.

During the encoding and decoding of the data in the frequency generation and detection modules displayed in Fig. 3 it is important not to produce floating point numbers, as this could lead to an unexpected behavior by the FPGA. If this interface is going to be used for an application whose sensor requires processing of floating-point numbers, a combination of microcontroller unit (MCU) and FPGA would be preferable, as the floating-point unit of the MCU could support the FPGA in coding and decoding.

The lower validation frequency of 1 Hz and 10 Hz was chosen due to visibility. The parameters in Table 1, only change the temporal behavior of the IF_curr_exp neuron. For other characteristics such as threshold, the default parameters were used.

The parameter set may have no biological plausibility, but it provides a 1:1 relation between spike input and output, so that the received data can be observed, since it is not possible to record meta data from the external device population. In addition, an experiment was carried out to ensure that the 1:1 relation is valid up to a frequency of 1 kHz. Thus, the results in Fig. 4 and Fig. 5 are also valid for frequencies above 10 Hz. Frequencies above 1 kHz should not be used, as the SpiNN-3 only reliably supports a real time simulation time step of 1 ms and therefore no calculation can be performed between two spikes above 1 kHz. The simulation step size of 1 ms is also the reason why the membrane potential in the left side of Fig. 4 only goes to approx. -63 mV. In the next simulation time step, the potential already reaches the threshold value and is therefore plotted again at -65 mV.

In comparison to [2] the presented interface has a considerably lower power consumption (242 mW vs. 1.65 W). This could be reduced even further if the stand-alone FPGA and not the development board were used. However, the higher power consumption is acceptable because, for example, the 7-segment display of the development board makes debugging the FPGA code much easier.

## 6      Conclusion

In this paper, a bidirectional interface for digital neuromorphic hardware is presented. This interface is able to establish all usual protocols as inputs and outputs for the communication between the SpNN-3 and any sensor technology. The main platform used is an FPGA, which allows parallel data processing. Furthermore, the interface is able to transmit all occurring data to the user via an integrated PC interface and to also receive data from the user. It is also demonstrated that this endeavor could be implemented with low power consumption due to the FPGA used, which also proposed this interface as a basis for mobile platforms.

**Acknowledgments.** The work described in this paper was supported by the Frankfurt University of Applied Sciences and the University of Cadiz. The authors acknowledge the valuable support of Dr. Andrew Rowley, Dr. Andrew Gait (University of Manchester) and Jens Rau (Frankfurt University of Applied Sciences).